# Rotation-induced significant modulation of near-field radiative heat transfer between hyperbolic nanoparticles

Yang Hu[1,2,3], Yasong Sun[1,2,*], Zhiheng Zheng[4], Jinlin Song[5], Kezhang Shi[6], and Xiaohu Wu[3,7,*]

[1] Basic Research Center, School of Power and Energy, Northwestern Polytechnical University, Xi'an 710064, Shaanxi, P.R. China

[2] Center of Computational Physics and Energy Science, Yangtze River Delta Research Institute of NPU, Northwestern Polytechnical University, Taicang 215400, Jiangsu, P. R. China

[3] Shandong Institute of Advanced Technology, Jinan 250100, Shandong, P.R. China

[4] School of Energy and Materials, Shanghai Polytechnic University, Shanghai, 201209, P. R. China

[5] School of Electrical and Information Engineering, Wuhan Institute of Technology, Wuhan 430025, Hubei, P. R. China

[6] Centre for Optical and Electromagnetic Research, State Key Laboratory of Modern Optical Instrumentation, Zhejiang University, Hangzhou, 310058, P. R. China

[7] Shanghai Engineering Research Center of Advanced Thermal Functional Materials (Shanghai Polytechnic University), Shanghai 201209, China

*Email: yssun@nwpu.edu.cn and xiaohu.wu@iat.cn

**Abstract**

Modulation of near-field radiative heat transfer (NFRHT) with rotated anisotropic hBN and α-MoO$_3$ nanoparticles (NPs) are investigated. The spectral heat power, total heat power and electric field energy density are calculated at different particle orientations. Numerical results show that the modulation factor of the α-MoO$_3$ NPs could be up to ~12000 with particle radius of 40 nm at a gap distance of 200 nm, which is ~ 10-fold larger than the state of the art. Excitation of localized hyperbolic phonon polaritons (LHPPs) in different particle orientations allow for the excellent modulation. The modulation factor of the hBN NPs is 5.9 due to the insensitivity of the LHPPs in the type II hyperbolic band to the particle orientations. This work not only provides a new method to modulate NFRHT in particle systems, but also paves the way to explore radiative heat transfer characteristics of anisotropic media.

**1. Introduction**

Near-field radiative heat transfer (NFRHT) between two objects can exceed the black body limit by several orders of magnitude at the nanoscale [1-4]. Compared to the far-field thermal radiation, near-field evanescent waves/modes enable photon tunneling and allow for colossal energy transfer [5-7], giving promising potential applications in thermophotovoltaics [8], thermal rectification [9], noncontact refrigeration [10], thermal transistors [11] and so on. Excitation of surface plasmon polaritons (SPPs), surface phonon polaritons (SPhPs), hyperbolic phonon polaritons (HPPs) or their coupling modes can support near-unity photon tunneling for further enhancement [12-17]. Numerous studies have shown the outstanding advantages of hyperbolic materials, such as artificial metamaterials [18,19], hexagonal boron nitride (hBN) [20-23], and α-MoO$_3$ biaxial crystal [24-26], providing potential applications in thermal management and manipulation.

Previous studies have demonstrated that NFRHT modulation could be achieved with phase-change materials [27,28], magnetic-optical materials [29] and other interesting methods. For instance, Biehs et al. proposed to modulate the NFRHT by introducing a rotation angle between two grating [30], inspiring many studies of thermal radiation modulation based on planar materials [31-36]. Wu et al. showed an obvious change of heat flux by changing the orientation of hBN optical axis [20], and the ratio of maximum value to minimum value in modulation, i.e., modulation contrast is 12.45 with a hBN-α-MoO$_3$ configuration [37]. Recently, NFRHT between nanoparticles (NPs) in the presence of a substrate of magneto-optical materials [38,39] and two-dimensional materials [40-43], has attracted extensive attention. One can modulate the radiative heat transfer between NPs [44-46] due to the thermal relay effect of substrate. This could be further controlled by applying an external magnetic field [40], changing the drift currents [42,43] or tilting the optical axis of the substrate [47]. So far, the modulation of NFRHT between two-body NPs without the presence of substrate has not been fully studied.

In this work, we study the rotation modulation of NFHRT between anisotropic NPs, such as hBN and α-MoO$_3$. The total heat power, spectral heat power and electric field energy density with different particle orientations are discussed. The results show that the excitation of LHPPs can significantly enhance the NFRHT and affect rotation modulation due to the resonance modes coupling or mismatching. The modulation effect (up to ~ 12000) of biaxial material α-MoO$_3$ is better than that of the uniaxial material hBN, and is 10-fold larger than the state of the art. In addition, by properly reducing the inter-particle distance and increasing the particle radius, the control effect can be improved.

## 2. Modeling and calculation

Figure 1 shows the schematic illustration of NFRHT between two anisotropic NPs at different particle orientations. Here, two particles are described as point sources with radius smaller than the skin depth of the electromagnetic field. Thus, the dipolar approximation works when the distance between two particles is three times larger than the radius [48,50].

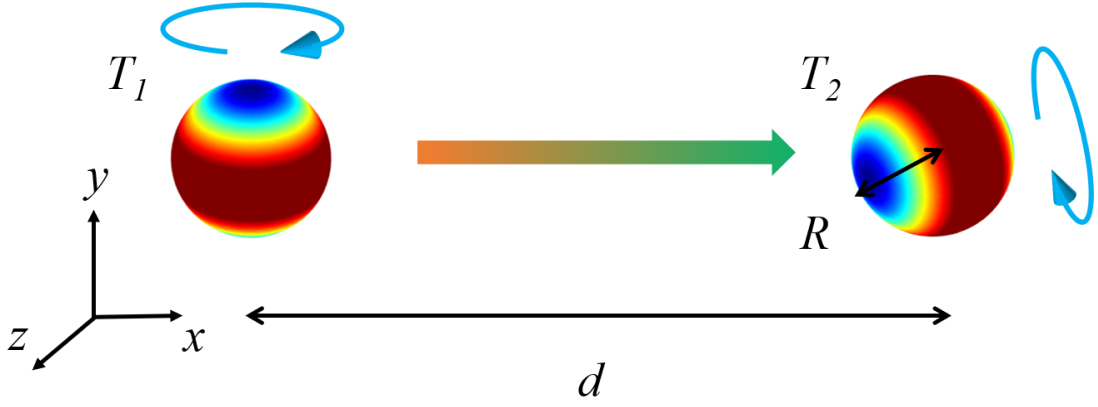

Fig. 1 Schematic of the NFRHT between two anisotropic NPs with radius *R* at a gap distance of *d*. The two particles are placed along the *x*-axis. The temperatures of the emitter and the receiver are $T_1$ and $T_2$, respectively. The different colors show anisotropic directional thermal radiation.

The optical responses of anisotropic materials are related to the orientation of their optical axis (OA). The dielectric functions of hBN and α-MoO$_3$ (as tensors) can be described by the following Lorentz equation [18,22]:

$$\varepsilon_m = \varepsilon_{\infty,m}\left(1 + \frac{\omega_{LO,m}^2 - \omega_{TO,m}^2}{\omega_{TO,m}^2 - \omega^2 + i\Gamma_m \omega}\right) \quad (1)$$

where $m = \perp, \parallel$ or $m = x, y, z$ for hBN or α-MoO$_3$, respectively. For hBN, $\omega_{TO,\perp}$ =2.58×10$^{14}$ rad/s, $\omega_{TO,\parallel}$=1.47×10$^{14}$ rad/s, $\omega_{LO,\perp}$=3.03×10$^{14}$ rad/s, $\omega_{LO,\parallel}$=1.56×10$^{14}$ rad/s, $\varepsilon_{\infty,\perp}$=4.87, $\varepsilon_{\infty,\parallel}$=2.95, $\Gamma_\perp$=9.42×10$^{11}$ rad/s, and $\Gamma_\parallel$=7.54×10$^{11}$ rad/s [20,21]. For α-MoO$_3$, $\varepsilon_{\infty,x}$=4, $\varepsilon_{\infty,y}$=5.2, $\varepsilon_{\infty,z}$=2.4, $\omega_{LO,x}$=1.83×10$^{14}$ rad/s, $\omega_{LO,y}$=1.60×10$^{14}$ rad/s, $\omega_{LO,z}$=1.89×10$^{14}$ rad/s, $\omega_{TO,x}$=1.54×10$^{14}$ rad/s, $\omega_{TO,y}$=1.02×10$^{14}$ rad/s, $\omega_{TO,z}$=1.80×10$^{14}$ rad/s, $\Gamma_x$=7.53×10$^{11}$ rad/s, $\Gamma_y$=7.53×10$^{11}$ rad/s, and $\Gamma_z$=3.76×10$^{11}$ rad/s [24,25].

When omitting the thermal emission to the background, the total heat power received by particle *i* can be written as [36]:

$$P_i = \int_0^\infty \frac{d\omega}{2\pi} P_{i,\omega} = 3\int_0^\infty \frac{d\omega}{2\pi}[\Theta(T_j) - \Theta(T_i)]\tau_{ji} \quad (2)$$

where $\Theta(T) = \hbar\omega/(\exp(\hbar\omega/k_B T)-1)$ is the mean energy of the Planck thermal harmonic oscillators without zero-point energy. $\hbar$ is the reduced Planck constant and $k_B$ is the Boltzmann constant. $\tau_{ji} = 4k_0^4 Tr[\boldsymbol{\chi}_i \mathbf{G}_{ij} \boldsymbol{\chi}_j \mathbf{G}_{ij}^+]/3$ is the transmission coefficients for the power exchanged between the NPs. $\boldsymbol{\chi}_i = \text{Im}[\boldsymbol{\alpha}_i] - k_0^3 |\boldsymbol{\alpha}_i|^2/(6\pi)$ is the response function of particle *i* when considering fluctuation-electrodynamic theory and avoiding unphysical effects [48]. For an anisotropic particle, based on the well-known Clausius-Mossoti form [49], when neglecting radiative correction, its polarizability as a function of permittivity tensor is frequency-dependent and orientation-dependent. The electric polarizabilities of the anisotropic NPs $\boldsymbol{\alpha}_i$ is given as

$$\boldsymbol{\alpha}_i = 4\pi R^3 \frac{\boldsymbol{\varepsilon}_i - \mathbf{1}\mathbf{I}}{\boldsymbol{\varepsilon}_i + 2\mathbf{I}} \tag{3}$$

**I** is the third order identity matrix. $\boldsymbol{\varepsilon}_1$ ($\boldsymbol{\varepsilon}_2$) is the permittivity tensor of emitter (receiver). $\mathbf{G}_{ij}$ in $\tau_{ji}$ denotes the dyadic Green tensor of the system, which can be written as

$$\mathbf{G}_{ij} = \mathbf{M}_{ij}^{-1} \mathbf{G}_{ij}^{(0)} \tag{4}$$

with $\mathbf{M}_{ij} = \mathbf{I} - k_0^4 \boldsymbol{\alpha}_i \boldsymbol{\alpha}_j \mathbf{G}_{ij}^{(0)} \mathbf{G}_{ij}^{(0)T}$ representing the multiple reflections between two particles. $\mathbf{G}_{ij}^{(0)}$ is vacuum contribution to the Green function, which is related to the location of particles [48]:

$$\mathbf{G}_{ij}^{(0)} = \mathbf{G}^{(0)}(\mathbf{r}_i, \mathbf{r}_j) = \frac{e^{ik_0 r_{ij}}}{4\pi r_{ij}} \left[ \left(1 + \frac{ik_0 r_{ij} - 1}{k_0^2 r_{ij}^2}\right) \mathbf{I} + \frac{3 - 3ik_0 r_{ij} - k_0^2 r_{ij}^2}{k_0^2 r_{ij}^2} \hat{\mathbf{r}}_{ij} \otimes \hat{\mathbf{r}}_{ij} \right] \tag{5}$$

in which $k_0$ is wave vector in vacuum and $r_{ij}=|\mathbf{r}_{ij}|$ is the magnitude of the vector linking position $\mathbf{r}_i$ and $\mathbf{r}_j$, $\hat{\mathbf{r}}_{ij} = \mathbf{r}_{ij}/r_{ij}$ and $\otimes$ denotes the exterior product.

In order to intuitively represent the NFRHT between two particles, spatial distribution of the radiated electric field energy density is introduced and expressed as [48]:

$$u_e(\mathbf{r_0},\omega) = \frac{2\varepsilon_0^2}{\pi\omega} \sum_i \Theta(\omega,T_i) Tr\left[Q_{r_0 i} \chi_i Q_{r_0 i}^*\right] \qquad (6)$$

in which $\mathbf{r}_0$ is a position outside the particle, $Q_{r_0 i} = \omega^2 \mu_0 G_{r_0 i}^{(0)} G_{r_0 i}$.

In the calculation, the temperatures of the emitter and receiver are $T_1=350$ K and $T_2=300$ K, the coordinate of emitter the receiver are $\mathbf{r}_1 = (a\ b\ c)$, and $\mathbf{r}_2 = (a+d\ b\ c)$, respectively. The distance between the emitter and receiver is $d=200$ nm, and the radius of both particles are $R=10$ nm unless otherwise specified.

### 3. Results and discussion

3.1 hBN NPs

Figure 2(a) shows the real part of permittivity components of hBN. The real parts of $\varepsilon_\perp$ and $\varepsilon_\parallel$ are negative within two frequency regions from $1.47\times10^{14}$ rad/s to $1.56\times10^{14}$ rad/s, and from $2.58\times10^{14}$ rad/s to $3.03\times10^{14}$ rad/s, respectively. These regions correspond to two Reststrahlen bands of hBN, in which the dispersion for electromagnetic wave propagation can exhibit the property of hyperbolicity. The existence of the hyperbolic bands greatly affects the NFRHT between hBN NPs. Fig. 2(b) shows the imaginary parts of polarizabilities of hBN, which are proportional to the absorptivity of NPs. When particles are surrounded by vacuum, the resonance frequencies can be obtained from the relationship $\boldsymbol{\varepsilon}_i + 2\mathbf{I} = 0$. Here, the resonance frequencies are $1.526\times10^{14}$ rad/s and $2.908\times10^{14}$ rad/s corresponding to $\alpha(\varepsilon_\parallel)$ and $\alpha(\varepsilon_\perp)$, respectively. At these frequencies, the excitation of LHPPs can significantly enhance NFHRT between two particles [43].

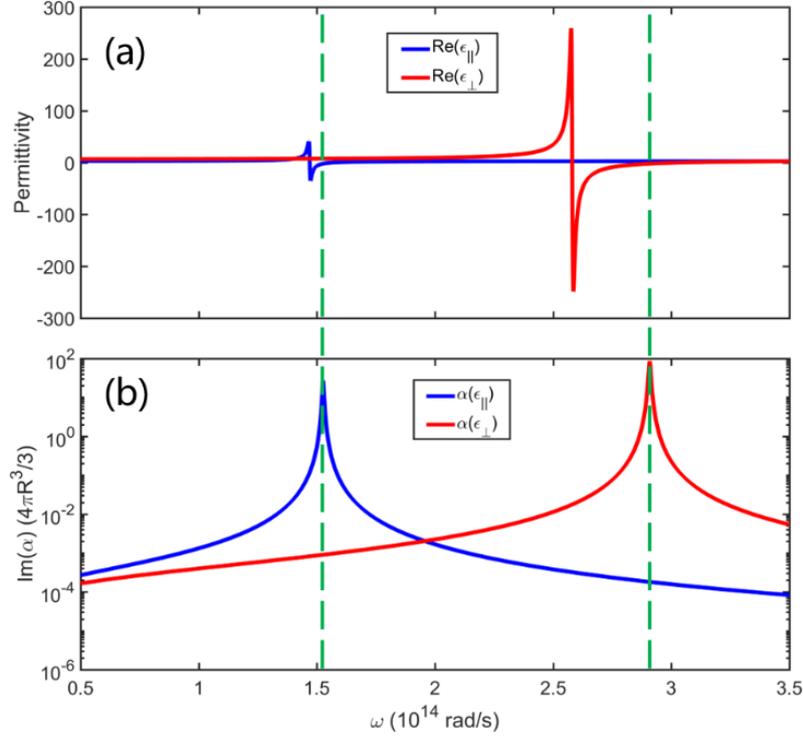

Fig. 2 (a) Real parts of permittivity components $\varepsilon_\parallel$ and $\varepsilon_\perp$ as well as (b) imaginary parts of polarizabilities $\alpha(\varepsilon_\parallel)$ and $\alpha(\varepsilon_\perp)$ normalized to the particle volume $4\pi R^3/3$.

For the uniaxial material hBN, the permittivity tensor varies with the direction of OA. When OA is along $x$-axis, $y$-axis, and $z$-axis, the permittivity tensor can be expressed as $\boldsymbol{\varepsilon}_i = \mathrm{diag}(\varepsilon_\parallel, \varepsilon_\perp, \varepsilon_\perp)$, $\boldsymbol{\varepsilon}_i = \mathrm{diag}(\varepsilon_\perp, \varepsilon_\parallel, \varepsilon_\perp)$ and $\boldsymbol{\varepsilon}_i = \mathrm{diag}(\varepsilon_\perp, \varepsilon_\perp, \varepsilon_\parallel)$, respectively. As shown in Fig. 3, the total heat power between the emitter and receiver with different OA can be divided into six cases. In general, the NFRHT of the hBN NPs are variable at different particle orientations. The total heat power with $\boldsymbol{\varepsilon}_1 = \mathrm{diag}(\varepsilon_\parallel, \varepsilon_\perp, \varepsilon_\perp)$ and $\boldsymbol{\varepsilon}_2 = \mathrm{diag}(\varepsilon_\perp, \varepsilon_\parallel, \varepsilon_\perp)$ is same as that with $\boldsymbol{\varepsilon}_1 = \mathrm{diag}(\varepsilon_\parallel, \varepsilon_\perp, \varepsilon_\perp)$ and $\boldsymbol{\varepsilon}_2 = \mathrm{diag}(\varepsilon_\perp, \varepsilon_\perp, \varepsilon_\parallel)$. That is because hBN is uniaxial and it shows isotropic in the plane perpendicular to OA. When the particles are placed along *x*-axis of coordinate system, it demonstrates rotation symmetry in *y-z* plane. This also accounts for the same results between [$\boldsymbol{\varepsilon}_1 = \text{diag}(\varepsilon_\perp, \varepsilon_\parallel, \varepsilon_\perp)$, $\boldsymbol{\varepsilon}_2 = \text{diag}(\varepsilon_\perp, \varepsilon_\parallel, \varepsilon_\perp)$] and [$\boldsymbol{\varepsilon}_1 = \text{diag}(\varepsilon_\perp, \varepsilon_\perp, \varepsilon_\parallel)$, $\boldsymbol{\varepsilon}_2 = \text{diag}(\varepsilon_\perp, \varepsilon_\perp, \varepsilon_\parallel)$]. Furthermore, the total heat power between two particles with the same orientations are always larger than that with dissimilar orientations. Maximum heat power ($P=1.210\times10^{-15}$ W) could be obtained with OA perpendicular to the *x*-axis, while it reduces to a minimum case ($P=2.051\times10^{-16}$ W) when the permittivity tensors change to $\boldsymbol{\varepsilon}_1 = \text{diag}(\varepsilon_\parallel, \varepsilon_\perp, \varepsilon_\perp)$ and $\boldsymbol{\varepsilon}_2 = \text{diag}(\varepsilon_\perp, \varepsilon_\parallel, \varepsilon_\perp)$.

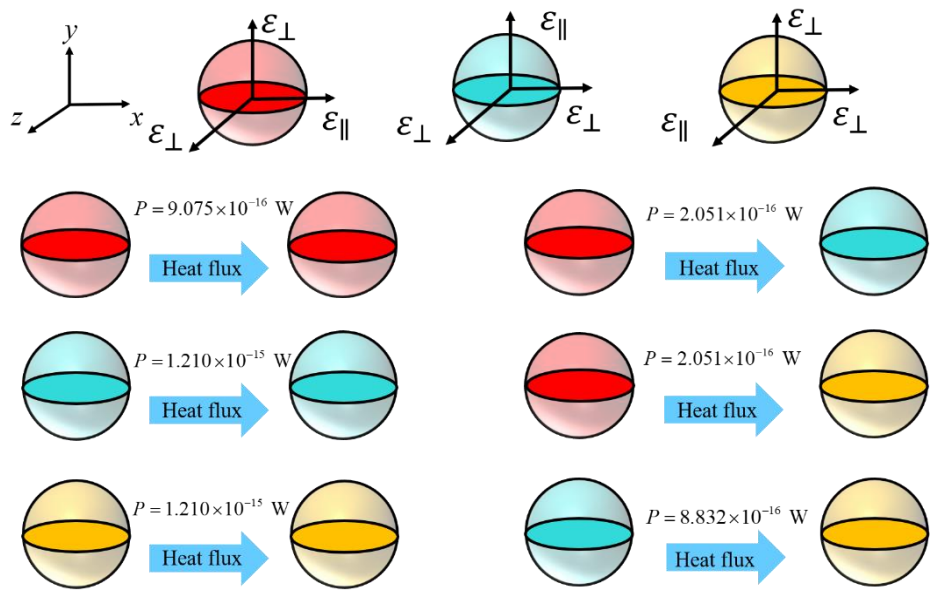

Fig. 3 Permittivity tensors and total heat power between particles at different orientations of hBN particles.

The spectral power with different orientations of the particles are plotted in Fig. 4(a). Two peaks attributed to the LHPP resonances with angular frequencies of $1.526\times10^{14}$ rad/s and $2.908\times10^{14}$ rad/s are observed, which are independent on the orientations. The intensity of the peaks at frequencies of $1.526 \times 10^{14}$ rad/s are always larger when the emitter and receiver have the same OA, which are variable sensitively with different orientations, indicating the modulation towards the total heat power. In contrast, those peaks at frequencies of $2.908 \times 10^{14}$ rad/s seem to be the same during the rotation process, which is quite different compared to the α-MoO$_3$ NPs in Fig. 10.

The rotation of particles from the case of minimum heat power to the case of maximum heat power are further considered. The total heat power as a function of rotation angle are calculated by changing the OA of the emitter in *x-y* plane (when the receiver remains unchanged). The inset in Fig. 4(b) shows the rotation process. As a result, the total heat power increases monotonically with respect to the rotation angle $\beta_1$. Note that $\beta_1 = 0.5\pi$ represents the case of $\boldsymbol{\varepsilon}_1 = \mathrm{diag}(\varepsilon_\perp, \varepsilon_\parallel, \varepsilon_\perp)$ and $\boldsymbol{\varepsilon}_2 = \mathrm{diag}(\varepsilon_\perp, \varepsilon_\parallel, \varepsilon_\perp)$, which performs the best as the matching LHPP modes of the identical emitter and receiver. The modulation factor (defined as the ratio of the maximum to the minimum) is 5.9 in this case.

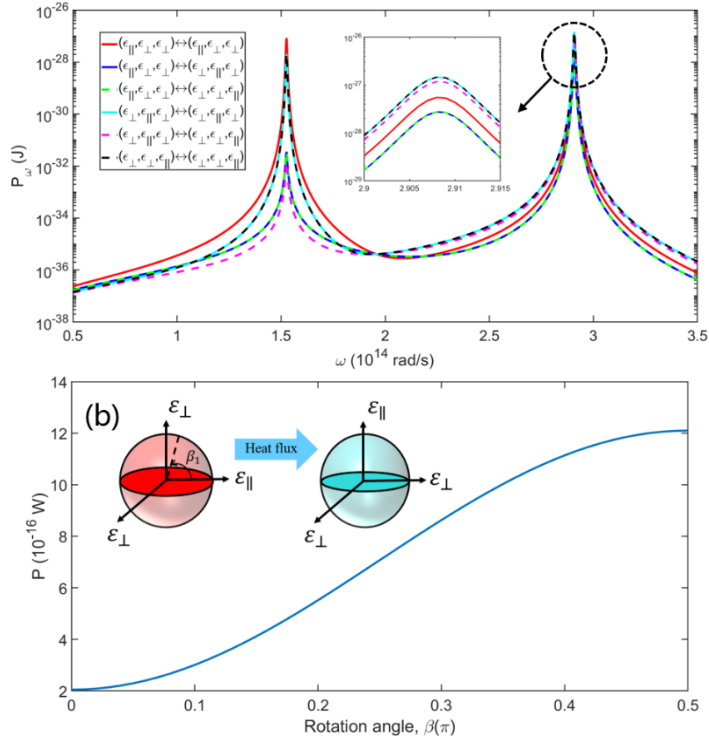

Fig. 4 (a) Spectral power between anisotropic hBN-NPs at different orientations varying with angular frequency. (b) Total heat power between the hBN-NPs with variable rotation angle of the emitter. $\beta_1 = 0$ represents the case with the permittivity tensors of $\boldsymbol{\varepsilon}_1 = \mathrm{diag}(\varepsilon_\parallel, \varepsilon_\perp, \varepsilon_\perp)$ and $\boldsymbol{\varepsilon}_2 = \mathrm{diag}(\varepsilon_\perp, \varepsilon_\parallel, \varepsilon_\perp)$, while $\beta_1 = 0.5\pi$ refers to $\boldsymbol{\varepsilon}_1 = \mathrm{diag}(\varepsilon_\perp, \varepsilon_\parallel, \varepsilon_\perp)$ and $\boldsymbol{\varepsilon}_2 = \mathrm{diag}(\varepsilon_\perp, \varepsilon_\parallel, \varepsilon_\perp)$.

To give an intuitive explanation of underlying physics, the spatial distributions of the radiated electric field energy density are plotted in Fig. 5. The angular frequency selected in this figure is the resonance frequency in type I hyperbolic band, viz., $\omega = 1.526 \times 10^{14}$ rad/s. At this frequency, there is $\varepsilon_\perp = 7.72 + 0.01i$ and $\varepsilon_\parallel = -1.98 + 0.33i$. The orientations of these two particles are identical in Fig. 5. The first, second and third rows in the figure represent the OA along the $x$, $y$ and $z$ directions, respectively. The first and the second columns represent the $x$-$y$ plane view and the $x$-$z$ plane view, respectively. As shown in Fig. 5(a), the energy density near the two particles is large and nonuniform around the particles. The energy density along *x* direction is greater than that in *y* direction as the permittivity in *x* direction is negative and close to -2, where the LHPPs can be excited. Furthermore, when two particles are placed in the same direction along the OA, the excitation effect can be strongly coupled and greatly enhances the NFRHT. It can be found from Fig. 5(a) and 5(b) that when OA is along *x*-axis and particles are placed along *x* direction, the *x-y* view is same as *x-z* view, that is due to hBN is a uniaxial material with the same components of permittivity in *y* and *z* direction. A similar phenomenon can be found in Fig. 5(b), when the permittivity is negative, the NFHRT is obviously enhanced. While in Fig. 5(e), when the permittivity is positive in *x* and *z* direction, there is no excitation of LHPPs, the energy density distribution is uniform around the particle, and the particles can be seen as isotropic in *x-z* plane. The phenomenon in Fig. 5(c) and (f) are similar as (e) and (b), respectively．The heat flux between the two particles remains the same when OA is along *y* and *z*.

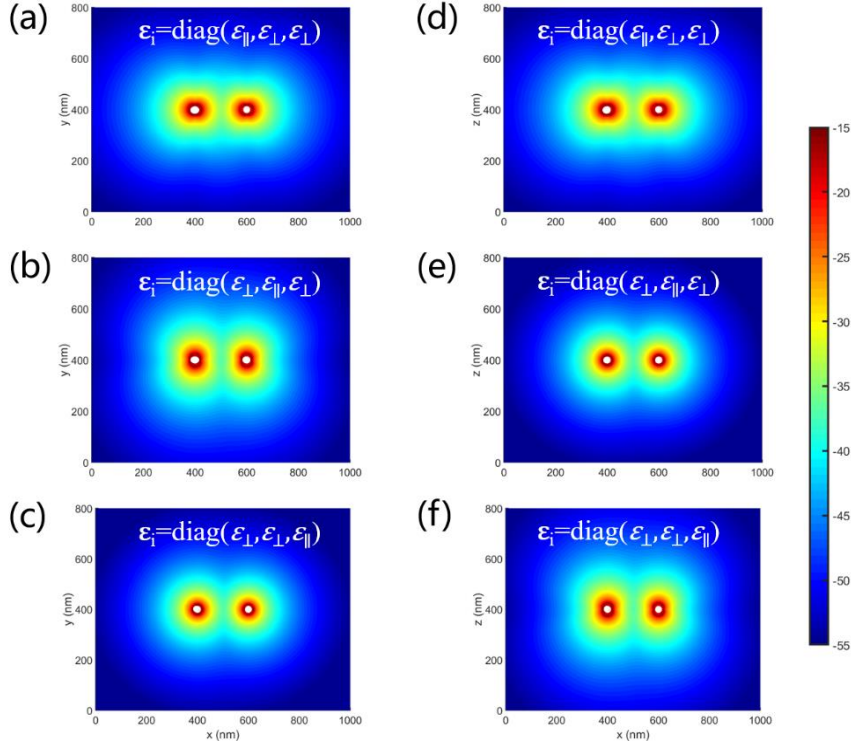

Fig. 5 The electric field energy density distribution at the frequency $1.526\times10^{14}$ rad/s for two identical hBN nanoparticles. $\mathbf{r}_1$ = (400, 400, 400) (nm), $\mathbf{r}_2$= (800, 400, 400) (nm). (a), (b) and (c) are *x-y* plane view when *z* is 400 nm, (d), (e) and (f) are *x-z* plane view when *y* is 400 nm. The permittivity tensors of particles are shown at the top of the figures. The white circles represent the particle position.

In order to explain the energy change of different particle orientations, we plot spatial distribution of the radiated electric field energy density with different particle orientations in Fig. 6. As illustrated in Fig. 6(a), when the OA of two particles are along *x* direction, energy is highly confined between the particles due to the strong coupling of the LHPPs. However, when the excitation direction is different, the LHPP modes are mismatching (as shown in Figs. 6(b), (c) and (d)), resulting in less total heat power between two particles.

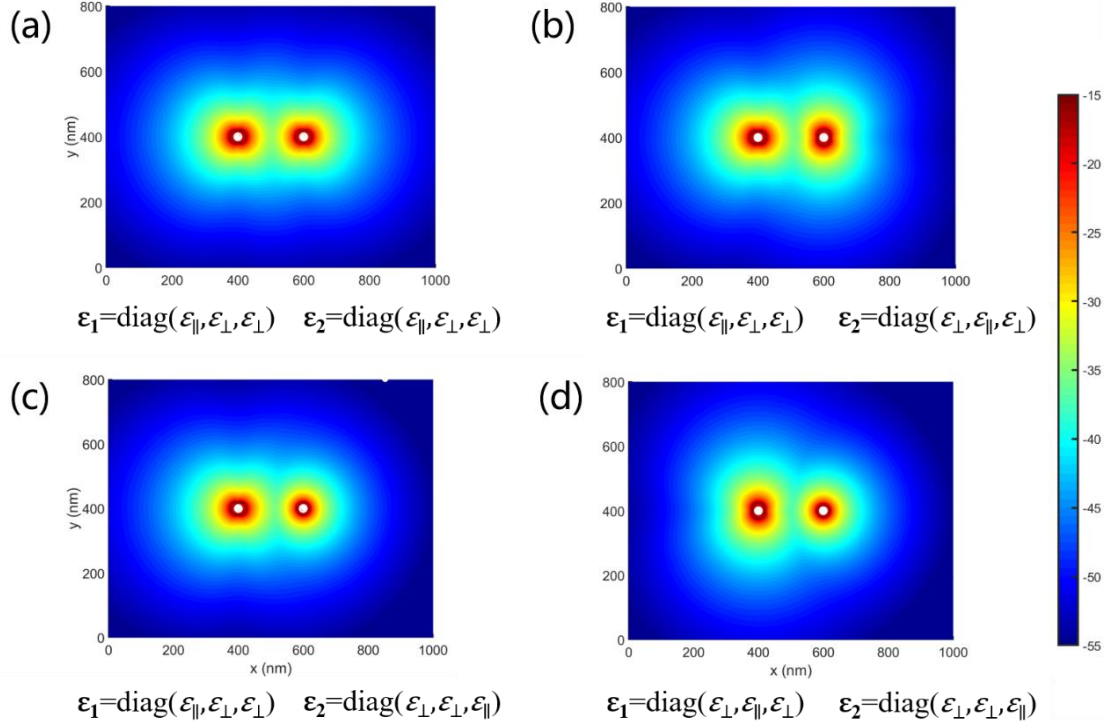

Fig. 6 The electric field energy density distribution at the frequency $1.526\times10^{14}$ rad/s for two hBN nanoparticles in different orientations in *x-y* plane view. The permittivity tensors of emitter and receiver are shown at the bottom of the figures.

Furthermore, the effects of rotating particles to the resonance frequency in type II hyperbolic band are discussed. The energy density of the electric field distributed in *x-y* plane is shown in Fig. 7, where the angular frequency is $2.908\times10^{14}$ rad/s $\varepsilon_\perp = -2.01+0.10i$, and $\varepsilon_\parallel = 2.82+0.0005i$. As the components of the permittivity tensor perpendicular to the OA are negative, the excitation of LHPPs could be observed. Compared with the type I hyperbolic band, LHPPs excitation here distribute in a larger space, and is insensitive to the particle orientations. The LHPP modes generated by the two particles can be better coupled in type II hyperbolic band. When two orientations of particles are different, excellent LHPPs resonance matching in type II hyperbolic band result in large spectral power compared with that within type I hyperbolic band.

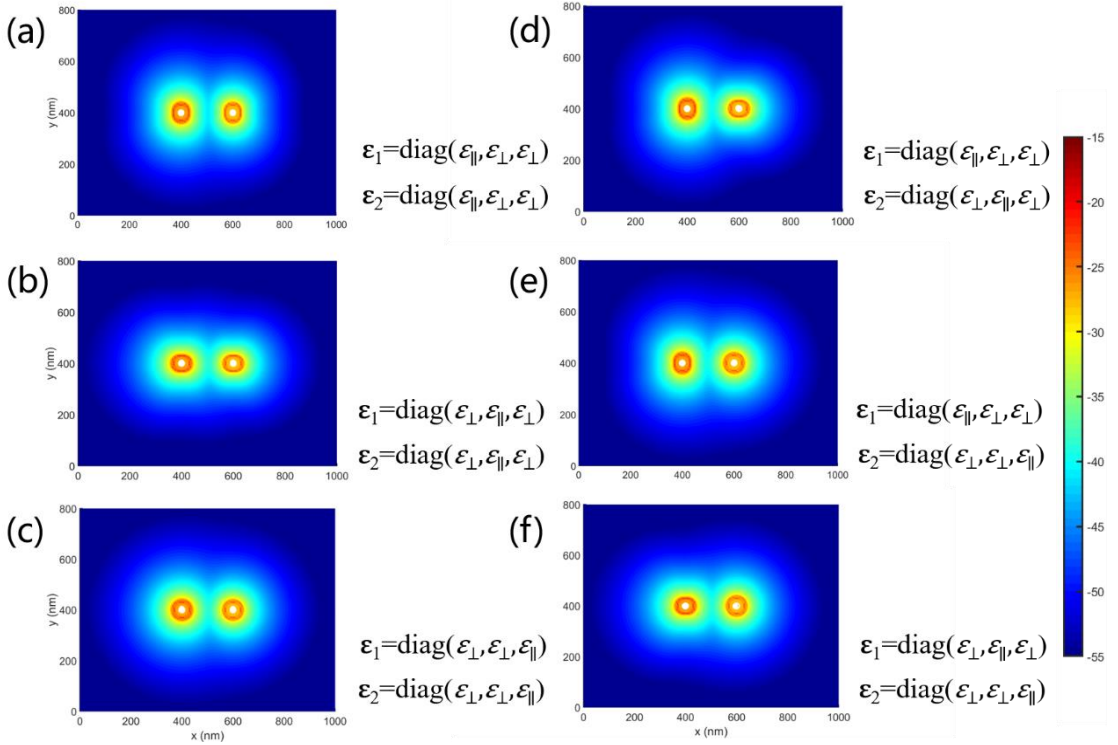

Fig. 7 The electric field energy density distribution at the frequency of $2.908\times10^{14}$ rad/s for two hBN nanoparticles in different orientations in *x-y* plane view. The permittivity tensors of emitter and receiver are shown at the right of the figure.

3.2 α-MoO$_3$ NPs

Compared to the hBN-NPs, the NFRHT between two biaxial α-MoO$_3$ particles are analyzed for larger modulation factor. Fig. 8(a) gives the real parts of permittivity components. As the biaxial material, α-MoO$_3$ features three crystalline directions along [0 1 0], [1 0 0], and [0 0 1], with respect to the permittivities of $\varepsilon_{zz}$, $\varepsilon_{xx}$, and $\varepsilon_{yy}$ [21]. The real parts of $\varepsilon_{xx}$, $\varepsilon_{yy}$ and $\varepsilon_{zz}$ are negative in three frequency regions ($1.027\times10^{14}$ rad/s, $1.604\times10^{14}$ rad/s), ($1.546\times10^{14}$ rad/s, $1.832\times10^{14}$ rad/s) and ($1.806\times10^{14}$ rad/s, $1.893\times10^{14}$ rad/s), respectively. These regions correspond to three Reststrahlen bands of α-MoO₃. Fig. 8(b) shows the imaginary parts of polarizabilities of α-MoO₃, where the three resonance frequencies of α-MoO₃ are $1.467\times10^{14}$ rad/s, $1.742\times10^{14}$ rad/s and $1.855\times10^{14}$ rad/s corresponding to $\alpha(\varepsilon_{yy})$, $\alpha(\varepsilon_{xx})$, and $\alpha(\varepsilon_{zz})$, respectively.

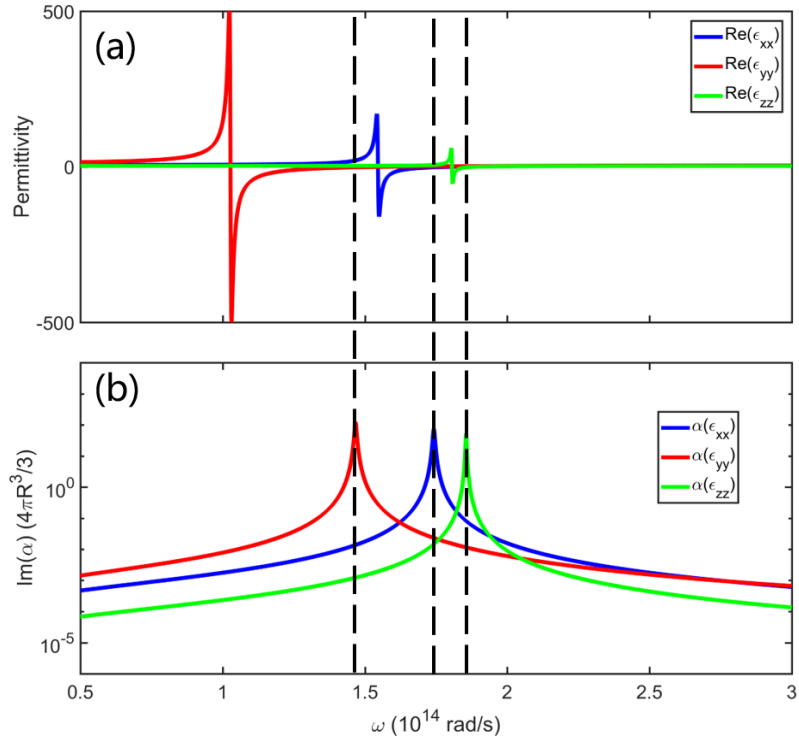

Fig. 8 (a) Real parts of permittivity components $\varepsilon_{xx}$, $\varepsilon_{yy}$ and $\varepsilon_{zz}$ as well as (b) imaginary parts of polarizabilities $\alpha(\varepsilon_{xx})$, $\alpha(\varepsilon_{yy})$ and $\alpha(\varepsilon_{zz})$ normalized to the particle volume $4\pi R^3/3$.

Fig. 9 shows the total heat powers between the α-MoO₃ NPs and their corresponding permittivity tensors. As a result, when the permittivity tensors of two particles are the same, the total heat power is distinctly larger. The maximum heat power ($P=1.214\times10^{-14}$ W) could be obtained with $\boldsymbol{\varepsilon}_1 = \mathrm{diag}(\varepsilon_{yy},\varepsilon_{zz},\varepsilon_{xx})$ and $\boldsymbol{\varepsilon}_2 = \mathrm{diag}(\varepsilon_{yy},\varepsilon_{zz},\varepsilon_{xx})$. Due to the mode mismatch, it significantly reduces to $P=2.306\times10^{-18}$ W when $\boldsymbol{\varepsilon}_1 = \text{diag}\left(\varepsilon_{zz},\varepsilon_{xx},\varepsilon_{yy}\right)$ and $\boldsymbol{\varepsilon}_2 = \text{diag}\left(\varepsilon_{yy},\varepsilon_{zz},\varepsilon_{xx}\right)$. The modulation factor reaches ~ 5264 far beyond the case of the hBN-NPs.

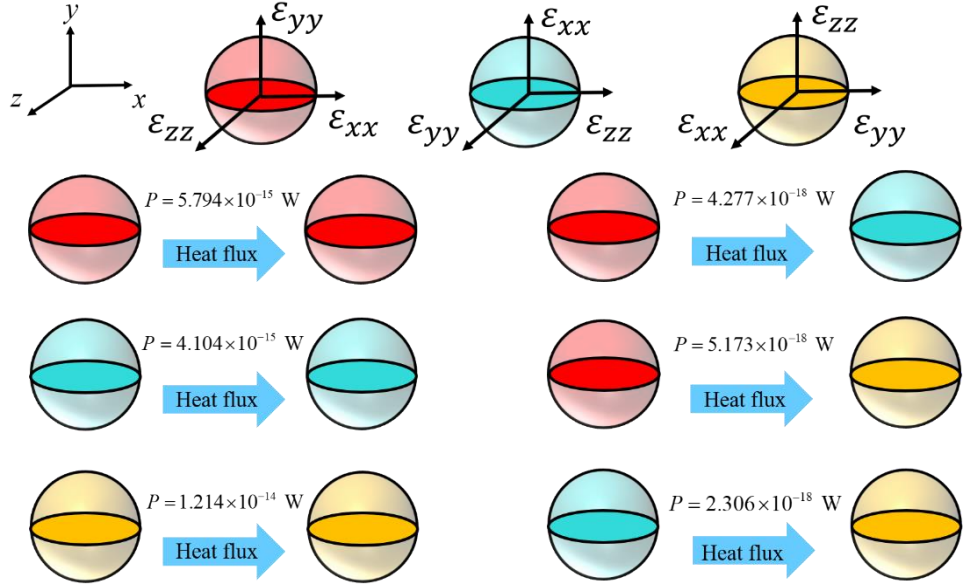

Fig. 9 Permittivity tensors and the total heat power between particles at different orientations of α-MoO$_3$ nanoparticles.

In order to further explore the characteristics of rotation modulation, Fig. 10 shows the spectral heat power at different particles orientations as a function of angular frequency. The minimum case shows a smaller spectral heat power within all frequencies (black-dashed line), which is different from that of the hBN-NPs in Fig.4a. For the maximum case, due to the LHPP modes matching, the spectral heat power is significantly improved within the three Reststrahlen bands (cyan-dashed line), account for the best performance on the total heat power in Fig. 9.

To clearly reveal rotation mechanism, the excitation of LHPP is discussed. LHPPs are excited at the resonance frequencies of $1.467\times10^{14}$ rad/s, $1.742\times10^{14}$ rad/s and $1.855\times10^{14}$ rad/s. At frequency of $1.467\times10^{14}$ rad/s, there is $\varepsilon_{xx}=20.29+0.76i$, $\varepsilon_{yy}=-2.00+0.01i$, and $\varepsilon_{zz}=3.12+0.004i$. The real parts of the two components in the permittivity tensor is positive, and the other is negative (close to -2). When real part of the permittivity is negative, LHPPs are excited. The excitation phenomenon at this frequency is similar to type I hyperbolic band of hBN. When the resonance frequency is $1.742\times10^{14}$ rad/s or $1.855\times10^{14}$ rad/s, the real part of two components in the permittivity tensor is negative and the excitation phenomenon of LHPPs are similar to $1.467\times10^{14}$ rad/s. LHPPs of α-MoO$_3$ NPs are confined in a narrow range in space at the resonance frequencies, resulting in efficient coupling or mismatching of the LHPP modes between two particles. Hence, the peak values vary sensitively in different orientations. Different from hBN NPs, the intensity of three peaks here are sensitive to the particle orientations, indicating the larger modulation towards the total heat power. Thus, controlling the matching of the LHPPs between α-MoO$_3$ NPs allows for remarkable NFRHT modulation.

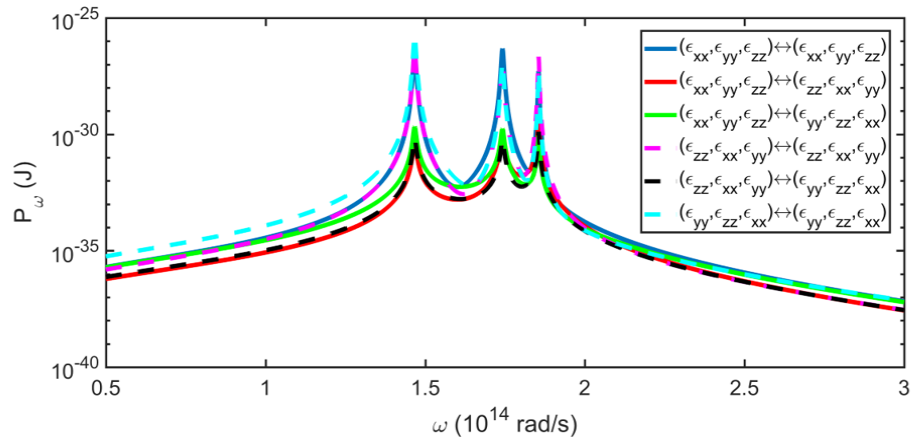

Fig. 10 Spectral heat power between two anisotropic α-MoO$_3$ NPs at different orientations.

Fig. 11 illustrates the rotation process between the maximum and minimum cases. As the rotation process can be reflected in the change of permittivity tensor of emitter, the receiver is modelling with $\boldsymbol{\varepsilon}_2 = \text{diag}(\varepsilon_{zz}, \varepsilon_{yy}, \varepsilon_{xx})$ while the emitter is rotated from $\boldsymbol{\varepsilon}_1 = \text{diag}(\varepsilon_{zz}, \varepsilon_{xx}, \varepsilon_{yy})$ to $\boldsymbol{\varepsilon}_1 = \text{diag}(\varepsilon_{zz}, \varepsilon_{yy}, \varepsilon_{xx})$. The emitter was firstly rotated around the $x$-axis in coordinate system with the permittivity tensor varying from $\boldsymbol{\varepsilon}_1 = \text{diag}(\varepsilon_{zz}, \varepsilon_{xx}, \varepsilon_{yy})$ to $\boldsymbol{\varepsilon}_1 = \text{diag}(\varepsilon_{zz}, \varepsilon_{yy}, \varepsilon_{xx})$. Next, the emitter was rotated around the $z$-axis and the permittivity tensor changed from $\boldsymbol{\varepsilon}_1 = \text{diag}(\varepsilon_{zz}, \varepsilon_{yy}, \varepsilon_{xx})$ to $\boldsymbol{\varepsilon}_1 = \text{diag}(\varepsilon_{yy}, \varepsilon_{zz}, \varepsilon_{xx})$. The total heat power of these two steps is shown in Fig. 9.

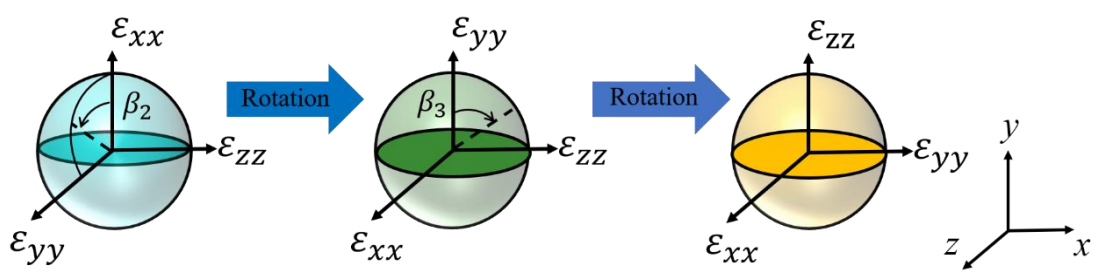

Fig. 11 Rotation processes between two particle orientations.

As shown in Fig. 12(a), the total heat power is monotone increasing with rotation angle $\beta_2$. The total heat power is $2.306 \times 10^{-18}$ W when the rotation angle $\beta_2$ is 0 (i.e., $\boldsymbol{\varepsilon}_1 = \text{diag}(\varepsilon_{zz}, \varepsilon_{xx}, \varepsilon_{yy})$), while it climbs up to $6.995 \times 10^{-16}$ W with rotation angle $\beta_2 = 0.5\pi$ (i.e., $\boldsymbol{\varepsilon}_1 = \text{diag}(\varepsilon_{zz}, \varepsilon_{yy}, \varepsilon_{xx})$). Hence, the total heat power increase sharply and the modulation factor reaches 303.3. Furthermore, when the emitter rotates along the $z$-axis and finally reaches $\beta_3 = 0.5\pi$, the total heat power reaches a maximum value ($1.214 \times 10^{-14}$ W), allowing for a remarkable modulation factor of ~5264. So far, the factor is ~ 5.36 for two hBN films [36,37], ~ 10 for two twisted grating [30], and ~ 1500 for NPs assisted by rotating substrate [47]. Hence, the giant modulation effect in this work is greater than any other previous studies.

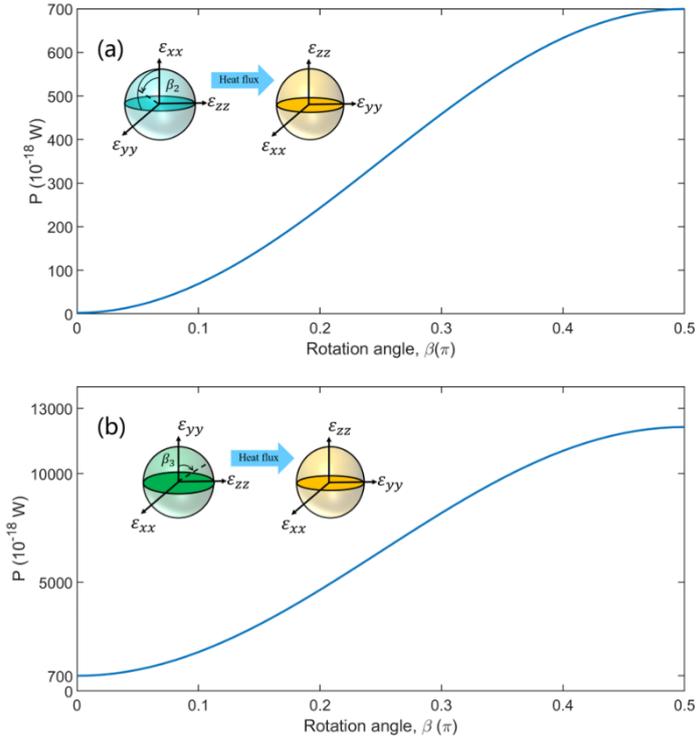

Fig. 12 Total heat power varying with rotation angle (a) rotate around *x*-axis, (b) rotate around *z*-axis.

The total heat power and modulation factor with variable distance and radius are also investigated. As shown in Fig. 13(a), both the total heat power and modulation factor decrease monotonically with distance. When the inter-particle distance is small, the total heat power decreases sharply with the distance, where *P* is approximately proportional to $d^{-6}$. The modulation factor increases remarkably with distance less than 70 nm, while is almost unchanged when distance is larger than 100 nm. Therefore, properly reducing the inter-particle distance can enhance the rotation modulation effect. Fig. 13(b) shows that the total heat power increase monotonically with radius, and when the particle radius is large, the growth trend is more intense. which is approximately proportional to $R^3$. The modulation factor increases monotonically with radius where a rapid increase could be observed after $R = 35$ nm. For example, the modulation factor reaches 12000 with $R = 40$ nm at a gap distance of 200 nm.

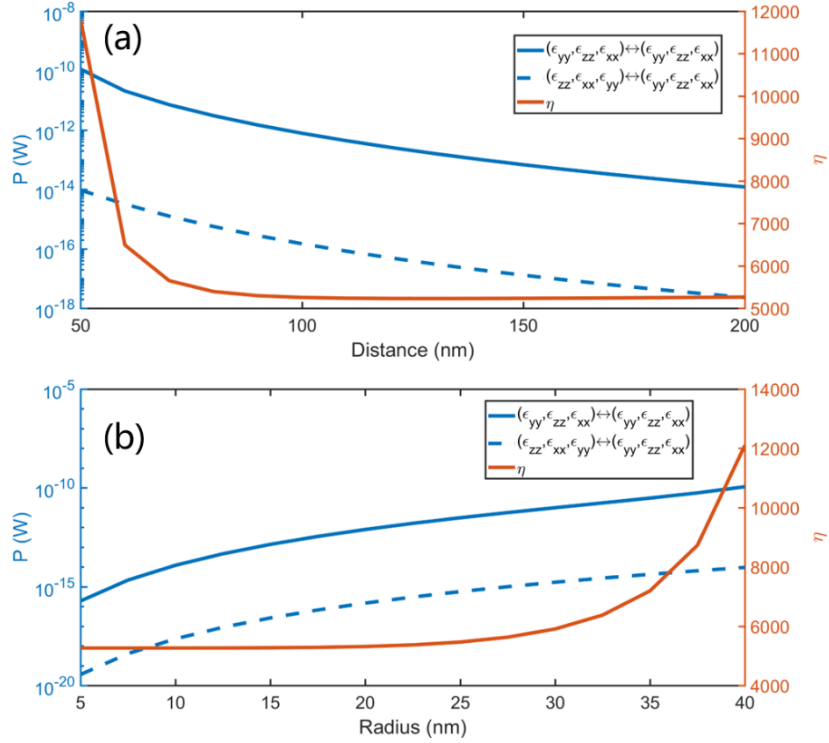

Fig. 13 Total heat power and modulation factor with variable distance and radius (a) $R$=10 nm (b) $d$=200 nm.

## 4. Conclusions

In summary, we have investigated the modulation of NFHRT by rotating anisotropic NPs of hBN and α-MoO$_3$. The total heat power of the NPs with different orientations are discussed. The spectral heat power and electric field energy density indicates that the excitation of LHPPs can greatly manipulate the NFHRT. The modulation factor of the α-MoO$_3$ NPs is found to be 12000 with $R = 40$ nm at a gap distance of 200 nm, which is much larger than previous works. The modulation factor could be optimized with smaller gap distance and larger radius of the NPs. This work may pave a new way to modulate NFHRT between NPs.

**Acknowledgments**

This work supported by the National Natural Science Foundation of China (Grant Nos. 52106099, and 51976173), the Shandong Provincial Natural Science Foundation (Grant No. ZR2020LLZ004), and the Jiangsu Provincial Natural Science Foundation (Grant No. BK20201204).